\documentstyle[pra,aps]{revtex}
\begin{document}
\draft
\title{Reduction criterion of separability and limits for a class of
protocols of entanglement distillation}

\author{Micha\l{} Horodecki \cite{poczta1}}

\address{Institute of Theoretical Physics and Astrophysics\\
 University of Gda\'nsk, 80--952 Gda\'nsk, Poland}

\author{Pawe\l{} Horodecki \cite{poczta2}}

\address{Faculty of Applied Physics and Mathematics\\
Technical University of Gda\'nsk, 80--952 Gda\'nsk, Poland}

\maketitle

\begin{abstract}
We analyse the problem of distillation of entanglement of mixed states in
higher dimensional compound systems.
Employing the positive maps method [M. Horodecki {\it et al.},
Phys. Lett. A {\bf 223} 1 (1996)] we introduce and analyse a criterion of
separability which relates the {\it structures} of the total density
matrix and its reductions.
We show that any state violating the criterion can be distilled by
suitable generalization of the two-qubit protocol which distills any
inseparable two-qubit state. Conversely, all the states which can be distilled
by such a protocol must violate the criterion. The proof involves construction
of the family of states which are invariant under transformation
$\varrho\rightarrow U\otimes U^*\varrho U^\dagger\otimes U^{*\dagger}$
where $U$ is a unitary transformation and star denotes complex conjugation.
The states are related to the depolarizing channel generalized to non-binary
case.
\end{abstract}


\section{Introduction}

Quatum entanglement \cite{EPR} appears to be one of the most
astonishing quantum phenomena. The new possibilities of applications
of the extremely strong quantum
correlations exhibiting by entangled states are being still discovered
\cite{Ekert,geste,Bennett_tel,huge,compl}.
Some of the theoretically predicted effects like
teleportation \cite{Bennett_tel} or quantum dense coding \cite{geste}
have been already realized experimentally \cite{geste_exp,Zeilinger}.
Most of those effects involves the paradigmatic entangled state which is the
singlet state of pair of spin-$1\over2$ particles \cite{Bohm}
\begin{equation}
\psi_s={1\over\sqrt2} (|\uparrow\rangle
|\downarrow\rangle-
|\downarrow\rangle
|\uparrow\rangle)
\label{singlet}
\end{equation}
This state cannot be reduced to direct product by any
transformation of the bases pertaining to each one of the particles.
Unfortunately, in practice we do not have singlet states in laboratory as
they evolve to mixed states due to uncontrolled interaction with environment.
However, the resulting mixtures may still contain some residual entanglement.
To be able to benefit the entanglement we must bring it to the singlet form
by means of local quantum (LQ) operations and classical communication (CC)
between the parties (typically Alice and Bob) sharing the pairs in
mixed state \cite{Bennett1}. Such process is called {\it purification}
of entanglement or {\it distillation}. Now, the fundamental question is

\noindent I. {\it Which mixed states can be distilled?}

To attempt to answer this question note that
the notion of entanglement can be naturally extended into mixed states
\cite{mixed}. Namely
we say that a mixed state $\varrho_{AB}$ acting on
a Hilbert space ${\cal H}={\cal H}_A\otimes {\cal H}_B$ is
inseparable (or entangled) if it
cannot be written in the form \cite{Werner}
\begin{equation}
\varrho=\sum_ip_i\varrho_A^i\otimes\varrho_B^i.
\label{separ}
\end{equation}
If instead, the state can be written in this way, we call it separable
(disentangled). Now, it is obvious that separable states cannot be distilled.
Indeed, LQ+CC operations cannot bring the separable state into inseparable
one, so that the final product cannot be the singlet state which is
manifestly inseparable. Then we may ask the following question

\noindent II. {\it Can any inseparable state be distilled?}

To answer this question, two kinds of effort had to be made.
First, given a density matrix, one did not have a way to check whether it is
separable or not. In other words, there was a problem of operational
characterization of separable (inseparable) states. The first attempt
to solve the problem was seeking necessary conditions
\cite{konieczne} of separability such as
criterion of violating of Bell's inequalities \cite{bell} (as one knows,
the separable states do not violate the inequalities), or the set
of entropic inequalities \cite{red,renyi,inf}. The very important step is due to
Peres \cite{Peres} who noted that separable states if partially
transposed remain positive.
Then by applying the machinery of positive maps formalism the Peres
condition has been shown \cite{sep} to be equivalent to separability
for $2\times2$ and $3\times2$ \cite{qubit} systems. For higher dimensions
explicit examples of inseparable states which do not violate Peres
condition have been constructed in Ref. \cite{Paw}.

To answer the question II, apart from the problem of characterization of
inseparable states, one needed to  investigate the protocols of distillation.
The original method of distillation introduced by Bennett {\it et al.}
\cite{Bennett1}
for  $2\times2$
systems (two-qubit ones) allows to distill  a state if and only if its
fully entangled
fraction $f$ is greater than $1\over2$. The quantity $f$ is defined as
\begin{equation}
f=\sup_{\psi}\langle\psi|\varrho|\psi\rangle,
\end{equation}
where the supremum is taken over all vectors $\psi$ which are of the form
$U_A\otimes U_B \psi_s$, where $U_A, U_B$ are unitary transformations and
$\psi_s$ is given by eq. (\ref{singlet}).
Another method was local filtering considered by Gisin \cite{Gisin}. He noted that some
states that  initially did not violate Bell's inequalities would do it if
subjected to local filtering \cite{Gisin}. This method does not lead to production of
singlet states from mixed ones. It was also considered by Bennett {et al.}
\cite{conc} in the context of
converting  pure non-maximally entangled states into singlet ones
(they call it
Procrustean method). In Ref. \cite{pur} the two protocols were put together and,
by use of the mentioned characterization of the inseparable $2\times2$
systems,
it was  shown that the question II has positive answer in the case of those
systems. The result can be easily extended to cover the $3\times2$ systems.
Now one could expect that all the inseparable states can be distilled and the
proof of that would be only question of time. Quite unexpectedly, a recent
result \cite{distbis} showed that it is not so. Namely, it turned out
that the states which do not violate the Peres criterion cannot be distilled.
Then, according to Ref. \cite{Paw} there are examples of mixed states
the entanglement of which cannot be brought to the singlet form!
Consequently, the answer to the question II is negative, and we should 
reformulate it as follows

\noindent III.  {\it Can all the states violating Peres condition be
distilled?}

The answer to this question is at present unknown.
The purpose of the present  paper is to contribute to solution of the problem.
We introduce separability condition based on
some positive map. The condition is
equivalent to separability for $2\times2$ (and $2\times3$)
systems. Moreover, it has the property
that any state (on an arbitrary $N\times N$ system i.\ e.\ consisted of
two N-level systems) which violates it, can be
effectively distilled by the suitable generalization of the  protocol
given in Ref. \cite{pur}.
The converse also holds: the only states which can be distilled by such a
kind of protocols, necessarily violate the criterion. Thus we obtain
limits of the use of the considered class of protocols.
One of the essential steps is determining the family of the states
which are invariant product unitary transformation of the form $U\otimes U^*$
where the star denotes complex conjugation. The family is connected
with the natural generalization of the quantum depolarizing channel to higher
dimensions.

We believe that the states violating the reduction criterion have  analogous
properties to the inseparable two-qubit states, so  that many distillation
methods introduced in the two-qubit case, if suitably generalized,
will work for the considered states.  In contrast, the inseparable states
satisfying the reduction criterion are supposed to exhibit oddities that do
not occur in the two-qubit case.


This paper is organized in the following way. In sec. \ref{CP}
we outline the method of investigation of inseparability by means of positive
maps. In sec. \ref{kryterium}
we present a separability criterion based on some positive map.
In particular, we show that it constitutes the necessary and sufficient
condition for separability for $2\times2$ and $2\times3$ systems and
is weaker than the Peres criterion for higher dimensions.
In sec. \ref{entropic}
we discuss it in the context of the entropic criteria relating the
density matrix of the system to its reductions.
In sec. \ref{rodzina} we derive the family of the
states which are invariant under random action of  $U\otimes U^*$
transformations. The family is connected with the depolarizing channel
generalized to higher dimensions. Subsequently, in sec. \ref{protokol},
we utilize the
results of the previous section to show that any state violating the
introduced reduction criterion can be distilled to the singlet form.
It is done via generalized XOR operation and
$U\otimes U^*$ twirling. We also point out that the criterion
determines the scope of use of a class of distillation protocols, namely
the ones consisting of two steps: one-side,
single-pair filtering and the procedure which cannot distill the states with
fully entangled fraction less than $1\over N$. In section \ref{ilustracja}
we illustrate the results by means of some examples.

\section{Positive maps, completely positive maps and inseparability}
\label{CP}

In quantum formalism the state of physical system is  represented by
density matrix, i.e., positive operator of unit trace. Positivity means
that the matrix is Hermitian and all its eigenvalues are nonnegative (if
an operator $\sigma$ is positive we write  $\sigma\geq0$). This assures
that diagonal elements of density matrix written in any basis are
nonnegative hence can be interpreted as probabilities of events. Thus to
describe the change of state due to physical process, we certainly need a
(linear) map which preserve positivity of operators (i.e. which
maps positive operators onto positive ones)
\begin{equation}
\sigma\geq 0 \Rightarrow\Lambda(\sigma)\geq0.
\end{equation}
Such maps are called {\it positive}  ones. However, it has been recognized
\cite{Lindblad} that the above condition is not sufficient
for a given map to describe a physical process. To see it imagine that we have
two systems $A$ and $B$ in some joint state $\varrho_{AB}$. Suppose
the systems
are spatially separated, so that each one  evolves separately and  the
evolution of the subsystems is given by $\Lambda_A$ and $\Lambda_B$.
Then the total evolution is described by
the map $\Lambda=\Lambda_A\otimes\Lambda_B$. Of course the
operator $\varrho_{out}=\Lambda(\varrho_{AB})$
describing the state after evolution must be still positive. It leads us to
another, very strong condition: the tensor multiplication of the  maps
describing the  physical processes must be still positive map.
The latter condition is called {\it complete positivity}.
In fact it appears that for a given map $\Lambda$ it suffices to have that
$\Lambda\otimes I_N$ is a positive map for each natural $N$, where $I_N: M_N\rightarrow
M_N$
denotes identity map acting on matrices $N\times N$ (i.e. the matrices
with N rows and N columns). This serves for definition of completely
positive (CP) map \cite{Lindblad}. For finite-dimensional systems
even weaker condition is sufficient (see Appendix). Finally, one can
distinguish an important subfamily of the CP maps which preserve trace i.e.
for which ${\rm Tr}\Lambda(\sigma)={\rm Tr} \sigma$.

In contrast with such a general and slightly abstract approach, one can
start by realizing what basic processes are allowed by quantum formalism.
There are the following ones
\begin{itemize}
\item[(i)] $\varrho\rightarrow \varrho\otimes\varrho'$ (adding a system
in state $\varrho'$)
\item[(ii)] $\varrho \rightarrow U\varrho U^\dagger$ (unitary transformation)
\item[(iii)] $\varrho_{AB}\rightarrow Tr_B\varrho_{AB}$ (discarding the system
- partial trace)
\end{itemize}
One can argue that any map describing physical process should allow to be
written by means of the above three maps \cite{Lindblad}.
In fact, it appears that comparison of  the  two approaches leads to
the satisfactory result: any trace-preserving CP map can be composed of the
above maps and, of course, all the three maps are trace-preserving
CP ones. If we supplement the three basic processes with the selection after
measurement, then we obtain the family of all CP maps.
Thus in the quantum formalism the most general physical process
the quantum state can undergo is described by CP map. In result, the
structure of the family of the  CP maps has been extensively investigated
\cite{Kraus,Lindblad} and
is at present well known. However, one knows that there exist positive
maps which are not CP ones. A common example is time reversal  operator
which acts as transposition of matrix in a given basis
\begin{equation}
(T \sigma)_{ij}=\sigma_{ji}
\end{equation}
To see that it is not CP, hence cannot describe a physical process
\cite{time,Woronowicz},
consider a two spin-$1\over2$ system in the singlet state given by
(\ref{singlet})
and suppose that one of the subsystems  is subjected to transposition while the
other one does not evolve. Then it is easy to see that the resulting operator
$A=(I\otimes T)(|\psi_s\rangle\langle\psi_s|)$ is not positive hence cannot
describe the state of physical system any longer. The time reversal is a
common example of the positive map which is not CP, however, one knows
only a few
other examples of such maps. Indeed, since the latter are of little
physical interest, their structure has  not been extensively investigated and
remains still obscure. However, recently we realized that they can be a
powerful tool in investigation of quantum inseparability of mixed states
\cite{sep}. To
see it, let us discuss in more detail the considerable fault of the
positive not-CP maps. The fault is that
there are states of compound systems (as the singlet state) which
are mapped by  $I\otimes \Lambda$ onto operators which are not positive. The
basic question is what features of the ``bad'' states cause the troubles?
To answer the question, recall that the singlet state is {\it entangled}
since it cannot be written as product of two state vectors describing
the subsystems. As mentioned in the introduction, the notion
of entanglement extends naturally to cover mixed states
(see formula (\ref{separ})). Now, let us 
note that there is no trouble with positive not-CP maps as long as
we deal with  separable states.
Indeed, in this case, if one of the systems is subjected by the
positive map then the resulting operator remains positive
\begin{equation}
(I\otimes\Lambda)\left(\sum_i p_i \varrho_A^i\otimes\varrho_B^i\right)=
\sum_i p_i \varrho_A^i\otimes\Lambda(\varrho_B^i)\geq 0
\label{nec}
\end{equation}
as $\Lambda(\varrho_i^B)\geq0$ due to positivity of $\Lambda$.
Thus if a positive map is not CP, this can be only recognized by means of
inseparable states.  In other words, it is just inseparability which
forced one to
remove some positive maps from the family of the maps corresponding the physical
processes. This suggests that the positive maps can be a particularly useful
tool for investigation of inseparability. Indeed, a theorem has been proved
 \cite{sep} stating that any state  is inseparable
if and only if there exists a positive map such that $(\Lambda\otimes
I)(\varrho)$ is not positive. In particular, if we have a positive
map which is not CP then it automatically provides a necessary condition of
separability which can be written as
\begin{equation}
(I\otimes \Lambda)(\varrho)\geq0
\label{cond}
\end{equation}
For a given map $\Lambda$, the map $I\otimes \Lambda$ will be further
denoted by $\tilde\Lambda$.


\section{Reduction criterion of separability}
\label{kryterium}

In this section we will utilize
the map  (acting on matrices $N\times N$)
of the form \cite{Kossakowski}
\begin{equation}
\Lambda(\sigma)=I {\rm Tr} \sigma-\sigma,
\label{map}
\end{equation}
where $I$ is identity matrix. It is easy to see that 
if $\sigma$ is positive then $I {\rm Tr} \sigma-\sigma$ also does,
hence $\Lambda$ is a
positive map. Writing the condition (\ref{cond}) explicitly for this particular
map we obtain \cite{ac}
\begin{equation}
\varrho_A\otimes I - \varrho\geq 0,
\label{kryt}
\end{equation}
where $\varrho_A={\rm Tr}_B\varrho $ is a reduction of the state  of interest.
Thus to use the criterion, one should find the reduction $\varrho_A$ and check
the eigenvalues of the operator $\varrho_A\otimes I- \varrho$.
Of course one can consider the dual criterion 
\begin{equation}
I\otimes \varrho_B - \varrho\geq 0.
\label{kryt2}
\end{equation}
As the two conditions relate the density matrix to its reductions
we will refer to  them taken jointly as to reduction criterion.

Let us now consider shortly the reduction criterion in the context
of the Peres one
\cite{Peres},
which writes explicitly
\begin{equation}
\varrho^{T_B}\geq  0.
\end{equation}
Here  $\varrho^{T_B}_{m\mu,n\nu}\equiv
\langle e_m\otimes f_\mu| \varrho^{T_B}|e _n \otimes f_\nu\rangle=
\varrho_{m\nu,n\mu}$ and $\{e_i\otimes f_j\}_{ij}$ is any product basis.
It is easy to see that both the criteria
are equivalent for the $2\times2$ (and $2\times3$) case. Indeed, the
map (\ref{map}) is in this case  of the form
$\Lambda(A)=(\sigma_y A\sigma_y)^T$ producing then equivalent criterion.

For higher dimensions, the map (\ref{map}) can be composed with
transposition and a completely positive map (see Appendix).
Hence, according to  \cite{sep}, if any state
violates the criterion (\ref{kryt}) then it must also violate
the Peres criterion \cite{Gin}.
Indeed, suppose that $\varrho$ satisfies the latter. Then
we have $\sigma\equiv \tilde T\varrho\geq0$, hence also for any CP map
$\Lambda_{CP}$
the operator $\tilde\Lambda_{CP}(\sigma)$ is positive. Consequently, if
a positive, but not CP map $\Lambda$ can be written as
$\Lambda=\Lambda_{CP}T$ (or equivalently $\Lambda=T\Lambda_{CP}$,
see Appendix) and a state satisfies Peres criterion, then it also satisfies
the criterion $\tilde\Lambda \varrho\geq0$ constituted by $\Lambda$.
Thus we see, that the reduction criterion is not stronger than the Peres one.

On the other hand, there exist states which satisfy the reduction criterion
but violate the Peres one. These are the Werner
states \cite{Werner} $W_N$ of $N\times N$
system given by
\begin{equation}
W_N=(N^3-N)^{-1}\left\{(N-\phi)I+
(N\phi-1) V\right\}
\label{werner}
\end{equation}
where $-1\leq\phi\leq 1$ and $V$ is
defined as $V \varphi\otimes\tilde\varphi=\tilde\varphi\otimes\varphi$.
The states are inseparable for $\phi<0$. For $2\times2$ system
the Werner states take a simple form \cite{Popescu}
\begin{equation}
W_2=(1-\alpha) {I\over 4} +
\alpha |\psi_s\rangle\langle\psi_s|,\quad -{1\over3}\leq \alpha\leq 1.
\label{Wer_Pop}
\end{equation}
being  mixtures of maximally chaotic state  and the singlet state.
It can be seen that all for $N\geq3$ inseparable Werner states 
violate partial transposition criterion satisfying the reduction one.
Indeed they have maximally mixed reductions and the norm less than
$1/N$, hence the inequality (\ref{kryt}) cannot be violated
(explicitly the reduction criterion for Werner states writes as $2-N\leq\phi\leq N$
which is satisfied for $N\geq3$)

The family of the Werner states
has an interesting property, namely they are the only states invariant under
any transformation of the form
\begin{equation}
\varrho\rightarrow U\otimes U\varrho U^\dagger\otimes U^\dagger,
\end{equation}
where $U$ is a unitary transformation (we say they are $U\otimes U$ invariant).
As we will see further, our criterion will lead  in a natural way
to distinguishing another family of states which are invariant under any
transformation of the form
\begin{equation}
\varrho\rightarrow U\otimes U^*\varrho U^\dagger\otimes U^{*\dagger}.
\end{equation}
where the star denotes complex conjugation of matrix elements of $U$
(analogously  we will call such states $U\otimes U^*$ invariant).
As it will be seen, the two families  are identical (up to a local
unitary transformation) for two-qubit case, but become distinct
for higher dimensions.

To summarise, in higher dimension the reduction criterion
is weaker than Peres one. The advantage of the present criterion is the
fact that, as it will be shown, all the states violating it can be distilled.
The latter result is compatible with \cite{distbis} where it is shown
that the states which can be distilled must violate the Peres criterion.

Finally,  there is a question whether  one could obtain stronger criterion by
applying the present one to a collection
$\varrho\otimes \cdots \otimes \varrho$
rather  than to state $\varrho$ of single pair (we will call it collective
application of the criterion). To answer the question it
is convenient to
introduce the following notation. If N parties share a number of M
N-tuples of particles, each one in state $\varrho_M$ then the joint
state $\varrho_1\otimes\cdots\otimes\varrho_M$  we will denote by
\begin{equation}
\left(\begin{array}{c}
\varrho_1\\
\varrho_2\\
\cdot\\ \cdot\\ \cdot   \\
\varrho_M
\end{array}\right)
\end{equation}
Consider first the Peres condition and apply it collectively. One can check
that \cite{Peres}
\begin{equation}
\tilde T\left({\varrho_1 \atop\varrho_2}\right)=
\left({\tilde T(\varrho_1) \atop\tilde T(\varrho_2)}\right).
\end{equation}
Hence, if the state $\varrho\otimes \varrho$ violates the criterion then
also $\varrho$ does, so that the collective application of Peres criterion
does not produce a stronger one.  Rains has proved \cite{Rains_priv}
that also in the case of the reduction criterion if the
state $\varrho_1\otimes \varrho_2$ of two pairs
violates it then the state of each pair separately also does.
Indeed, denoting the partial traces  of states $\varrho_1$   and
$\varrho_2$ over the systems B by $\tau_1$ and $\tau_2$ respectively, one
obtains
\begin{equation}
\tilde\Lambda\left({\varrho_1\atop\varrho_2}\right)=
\left({I\otimes\tau_1\atop I\otimes\tau_2}\right)-
\left({\varrho_1\atop\varrho_2}\right)
\end{equation}
hence
\begin{eqnarray}
&&\left({\tilde\Lambda(\varrho_1)\atop\tilde\Lambda(\varrho_2)}\right)=
\left({I\otimes\tau_1-\varrho_1\atop I\otimes\tau_2-\varrho_2}\right)=
\left({I\otimes \tau_1\atop I\otimes \tau_2}\right)+
\left({\varrho_1\atop\varrho_2}\right)-
\left({\varrho_1\atop I\otimes \tau_2}\right)-
\left({I\otimes \tau_1\atop \varrho_2}\right)=\\
&&=\tilde\Lambda\left({\varrho_1\atop\varrho_2}\right)+
2\left({\varrho_1\atop\varrho_2}\right)-
\left({\varrho_1\atop I\otimes \tau_2}\right)-
\left({I\otimes \tau_1\atop \varrho_2}\right).
\end{eqnarray}
This can be rewritten as
\begin{equation}
\tilde\Lambda\left({\varrho_1\atop\varrho_2}\right)=
\left({\tilde\Lambda(\varrho_1)\atop\tilde\Lambda(\varrho_2)}\right)+
\left({\varrho_1\atop\tilde\Lambda(\varrho_2)}\right)+
\left({\tilde\Lambda(\varrho_1)\atop\varrho_2}\right)
\end{equation}
Thus we have obtained that
\begin{equation}
\left[\tilde\Lambda(\varrho_1)\geq 0 \ {\rm and} \  \tilde \Lambda(\varrho_2)
\geq0\right]\Rightarrow
\tilde\Lambda\left({\varrho_1\atop\varrho_2}\right)\geq0
\end{equation}
which is the desired result.
\section{Reduction criterion and entropic ones}
\label{entropic}
Now it is interesting to discuss the criterion in the context of entropic
criteria which also exploit the relation between the total  system and its
subsystems. The first necessary condition of separability of this type was
constructed  by means of von Neumann entropies of the system and
subsystems \cite{red}.
The entropic criteria were then generalized by using  the quantum $S_\alpha$
Ren\'yi entropies  \cite{renyi}. They base on the
following inequalities
for conditional
entropies
\cite{renyi,inf}
\begin{equation}
S_\alpha(A|B)\geq 0,\quad S_\alpha(B|A)\geq 0
\end{equation}
with
\begin{equation}
S_\alpha(A|B)=S_\alpha(\varrho)-S_\alpha(\varrho_B),\quad
S_\alpha(B|A)=S_\alpha(\varrho)-S_\alpha(\varrho_A),
\end{equation}
where
\begin{equation}
S_\alpha={1\over 1-\alpha} \ln Tr\varrho^\alpha
\ \ \ {\rm for} \ \ \  1<\alpha<\infty,
\end{equation}
$S_1$ is the von Neumann entropy and
$S_\infty=-\ln\|\varrho\|$. It has been shown \cite{red,renyi,sep} that the
above
inequalities are satisfied by separable states for $\alpha=1,2$ and $\infty$.

The crucial difference is that they are {\it scalar} conditions being therefore
 weaker than the present criterion which relates the
{\it structures} of
the density matrix and its reductions rather than scalar functions of them.
This does not mean that the scalar criteria are useless. In fact, they can be
useful for  characterization of quantum channels. In particular, the von Neumann
conditional entropy has been recently used for definition of quantum
coherent information
\cite{Nielsen}. If, however, one is interested in  characterization of
separable (inseparable) states, the structural criteria are much more
convenient. In particular, note that  the reduction criterion is stronger
than the $\infty$
entropy inequality. The latter criterion says in fact that
for a separable state the largest eigenvalue of the
density matrix of the total system cannot exceed the one of any of the reduced
density matrices
\begin{equation}
\|\varrho\|\equiv \lambda_{\max}(\varrho)\leq\lambda_{\max}(\varrho_X)
\equiv\|\varrho_X\|, \quad X=A,B.
\label{infty}
\end{equation}
To see that this is implied by the conditions (\ref{kryt}),
(\ref{kryt2}), suppose that they are satisfied i.e.
$\varrho\leq \varrho_X\otimes I$ for X=A,B. Note that if
$0\leq \sigma_1\leq\sigma_2$ then we have also $0\leq\|\sigma_1\|\leq\|\sigma_2\|$.
Consequently, since $\varrho$ is positive and
$\|\varrho_X\|=\|I\otimes\varrho_X\|$, we immediately obtain
that $\|\varrho\|\leq\|\varrho\otimes I\|=\|\varrho_X\|$.
For states with maximally disordered subsystems the reduction criterion is
equivalent to the $\infty$ entropy inequality. Indeed, in this case
the smallest eigenvalue of the operator $\varrho_X\otimes I-\varrho$ is
equal to $\lambda={1\over N}-\|\varrho\|=\|\varrho_X\|-\|\varrho\|$, hence
both the criteria are satisfied or violated simultaneously.
Finally, the reduction criterion is essentially stronger than the
$\infty$ entropy inequalities, as the latter are not sufficient for
separability for two-qubit systems \cite{sep} while the reduction
criterion does, as shown above.

\section{ $U\otimes U^{*}$-invariant states.}
\label{rodzina}
In this section, applying the method used by Werner \cite{Werner}
we derive the family of $U\otimes U^*$ invariant
states i.e. the ones invariant under transformation of the form
$\varrho\rightarrow U\otimes U^*\varrho U^\dagger \otimes U^{*\dagger}$
(here $U$ is unitary transformation and $U^*$ - its complex conjugation).
Let us consider the Hermitian
operator $A$ which we want to
be $U\otimes U^*$ invariant. Let us write its matrix elements
in a product basis
\begin{equation}
\langle  mn|A|pq\rangle  \equiv \langle e_m \otimes
e_n|A|e_p  \otimes e_q\rangle.
\end{equation}
Imposing on A condition of $U\otimes U^*$ invariance with unitary operations $U$
taking some $|m_0\rangle $ to $-|m_0\rangle $ and leaving the other basis elements
unchanged we obtain that the only nonzero elements are of type
$\langle  mn|A|mn\rangle $, $\langle  mn|A|nm\rangle $ and $\langle  mm|A|nn\rangle $. Taking into account, again,
another set of unitary transformations,
each of the latter multiplying some single basis element
by imaginary unit $i$ and leaving the the remaining elements unchanged
we obtain immediately that all $\langle  mn|A|nm\rangle , m\neq n$ elements
must vanish. Considering now all two element permutations
of basis elements we obtain that the set of nonvanishing
matrix elements can be divided into three groups:
$\langle  mn|A|mn\rangle , m\neq n$, $\langle  mm|A|nn\rangle , m\neq n$ and $\langle  mm|A|mm\rangle $
with all elements in each of group equal.
Thus any $U\otimes U^*$ invariant Hermitian operator can be written
as $A=bB+cC+dD$ where: $B=\sum_{m\neq n}|mn\rangle \langle  mn|$, $C=\sum_{m\neq n}|mm\rangle \langle  nn|$,
$D=\sum_{m}|mm\rangle \langle  mm|$. Hermiticity of $A$ implies that parameters $b,c,d$
should be real. One can introduce the real unitary transformation of type
$U_1\otimes U_1=(\tilde{U}_2\oplus I_{N-2})\otimes(\tilde{U}_2\oplus I_{N-2})$
with
\begin{equation}
\tilde{U}_2=
\left[ \begin{array}{cc}
 cos\phi  & sin\phi \\
 -sin\phi &  cos\phi
\end{array} \right]
\end{equation}
acting on some two-dimensional subspace $H_1$ of $H$ and
$I_{N-2}$ being a projection on the space orthogonal to $H_1$.
It can be easily shown that the operator $D$ is not invariant
under $U_1\otimes U_1$ hence is not $U\otimes U^*$ invariant.
Thus parameter $d$ appears to be linearly dependent on $a$ and $b$.
Demanding, in addition ${\rm Tr}(A)=1$ we obtain that the
set of Hermitian $U\otimes U^*$ invariant operators with
unit trace are described by one real linear parameter.
On the other hand it can be checked immediately that
the family
\begin{equation}
\varrho_{\alpha}=(1-\alpha){I\over N^2}+\alpha P_+
\end{equation}
fulfils the above criteria.
Here $P_+=|\psi_+\rangle\langle\psi_+|$ with
\begin{equation}
\psi_+={1\over \sqrt{N}}\sum_{i=1}^{N}|i\rangle\otimes |i\rangle
\label{tryplet}
\end{equation}
is the generalized triplet state.
Indeed, the  identity operator is obviously $U\otimes U^*$ invariant and
for $P_+$ we obtain
\begin{equation}
U\otimes U^*P_+U^\dagger\otimes U^{*\dagger}=
I\otimes U^T U^*P_+I\otimes (U^T U^*)^\dagger=P_+
\end{equation}
where the property \cite{Jozsa} $A\otimes I \psi_+=I\otimes A^T \psi_+$
was used.
Imposing now the positivity condition, as we are interested in
$U\otimes U^*$ invariant {\it states}, we obtain the family
\begin{equation}
\varrho_{\alpha}=
(1-\alpha){I\over N^2}+\alpha P_+,\quad \ \rm{with} \quad
{-1\over N^2-1} \leq \alpha \leq 1
\label{family}
\label{alfa}
\end{equation}
Note that  $\varrho_\alpha$
can be viewed as Werner-Popescu state (\ref{Wer_Pop})
(mixture of singlet state and maximally chaotic state)
suitably generalized to higher dimensions.
The family can be parametrized by fidelity $F=\rm Tr \varrho P_+$ 
\begin{equation}
\varrho _{\alpha(F)} \equiv \varrho_F={N^2\over N^2-1}\left((F-1){I\over N^2}+(F-{1\over N^2})P_+\right),\quad 0\leq F\leq 1.
\label{prod}
\end{equation}
The above states are inseparable for fidelity greater than ${1 \over N}$
as they violate the criterion (\ref{kryt}). As $F$ is $U\otimes U^*$ invariant
parameter we obtain that for $F\leq {1 \over N}$
(resp. $\alpha\leq {1 \over N + 1}$) the states
can be reproduced by twirling
i. e. random $U \otimes U^{*}$ operation represented by the integral
\begin{equation}
\int U \otimes U^{*} ( \cdot ) U^{\dagger} \otimes {U^{*}}^{\dagger} dU
\end{equation}
performed on the proper {\it product}
pure state (here ${\rm d}\,U$ denotes
uniform probability distribution on unitary group $U(N)$ proportional to the
Haar measure).
This can be the state $P_{\phi}\otimes P_{\phi '}$ corresponding to the vectors
$\phi=|1\rangle $, $\phi ' =a|1\rangle +b|2\rangle $
with $F={|\langle \phi |\phi '  \rangle|^2 \over N}$. Thus the states
(\ref{prod}) (resp. \ref{alfa}) are inseparable {\it if and only if}
$F>{1 \over N}$ ($\alpha>{1 \over N+1}$).

The presented family defines the N-dimensional
{\it $\alpha$-depolarizing channel} which  completely randomises the input
state $\psi$ with probability $\alpha$ while leaves it undisturbed with
probability $1- \alpha$. Such a channel, in the case $N=2$, is being
extensively investigated at present \cite{huge,Shor}. As it will be shown,
the corresponding family of states (\ref{family}) resulting
from sending half of the state $P_+$ through the $(N,\alpha)$-depolarizing
channel can be distilled by means of LQ+CC operations 
if and only if $F>{1 \over N}$ ($\alpha>{1\over N+1}$). Then by using the results
relating quantum capacities and distillable entanglement \cite{huge} we obtain
that the considered channel supplemented by two way classical channel
has nonzero quantum capacity for this range of $\alpha$. This reproduces the
known result $F>{1 \over 2}$ ($\alpha>{1\over 3}$) for $N=2$ \cite{huge}.

\section{Distillation protocol}
\label{protokol}

Now our goal is to distill the states which violate the condition (\ref{kryt}).
The first stage  (filtering) \cite{Gisin} will be almost exactly the same as in
the protocol given in \cite{pur}.
For this purpose rewrite the condition (\ref{kryt}) in the form
\begin{equation}
\langle \psi|\varrho_A\otimes I- \varrho|\psi\rangle \geq0 \quad {\rm for\  any }\quad
\psi\in C^N\otimes C^N, \ ||\psi||=1,
\label{kryt4}
\end{equation}
or
\begin{equation}
{\rm Tr}\varrho P_\psi\leq {\rm Tr} \varrho_A\varrho_A^\psi
\label{in}
\end{equation}
where $P_{\psi}=|\psi\rangle \langle\psi|$ and $\varrho_A^\psi$
is reduced density
matrix	of $P_{\psi}$. Note that if we take $P_{\psi}$ being maximally entangled
states and maximize the left hand side over them, we will obtain the condition
for fully entangled fraction
\cite{Bennett1,huge} (generalized to higher dimensions)
\begin{equation}
f(\varrho)\equiv \max_{\Psi} Tr(\varrho P_{\Psi}).
\label{f}
\end{equation}
where the maximum is taken over all maximally entangled $\Psi$'s.
Namely we then have
\begin{equation}
f(\varrho)\leq {1\over N}
\end{equation}
for any separable $\varrho$.
Suppose now that a state $\varrho$ violates the condition (\ref{in}) for a
certain vector
\begin{equation}
|\psi\rangle=\sum_{m,n} a_{mn}|m\rangle\otimes |n\rangle
\end{equation}
Now, any such a vector can be produced from the triplet state $\psi_+$
given by (\ref{tryplet})
in the following way
\begin{equation}
|\psi\rangle =A\otimes I|\psi_+\rangle
\end{equation}
where $\langle m|A|n\rangle=\sqrt{N}a_{mn}$.
It can be checked that $AA^\dagger =N \varrho_A^{\psi}$.
Then, the  new state
\begin{equation}
\varrho'={A^\dagger\otimes I\varrho A\otimes I\over
{\rm Tr} (\varrho AA^\dagger \otimes I)}
\end{equation}
resulting from filtering $\varrho$ by means of one-side action
$A^\dagger\otimes I \varrho A\otimes I$ satisfies
\begin{equation}
{\rm Tr}  \varrho' P_+>{1\over N},
\label{fraction}
\end{equation}
Now, the problem is how to distill states with the property
(\ref{fraction}). For this purpose we need to generalize the protocol
\cite{Bennett1} used for two-qubit case. The first thing we need is the
generalized twirling procedure which would leave the state $P_+$ unchanged.
This, however, cannot be application of random bilateral unitary
transformation of
the form  $U\otimes U$ as there  is no $U\otimes U$ invariant pure state
in higher dimensions (this can be seen  directly from
the form of the Werner states (\ref{werner}))
As we have shown in sec. \ref{rodzina}
the suitable generalization we obtain applying
randomly transformation $U\otimes U^*$ where the star denotes complex
conjugation in any basis
(e.g. in the basis $|i\rangle$). From the results of the previous
section it follows that  for any $\varrho$
if ${\rm Tr}\varrho P_+=F$ then
\begin{equation}
\int U\otimes U^* \varrho U^\dagger \otimes U^{*\dagger}
{\rm d}\,U=\varrho_{\alpha}
\equiv (1-\alpha){I\over N}+\alpha P_+\quad \ \rm{with} \
\alpha={N^2F-1\over N^2-1},
\ \ 0 \geq F \geq 1
\end{equation}
i.e. after twirling we obtain state $\varrho_\alpha$ with the same fidelity
$F$ as the initial state. As it was shown in previous section,
the states $\varrho_\alpha$ are inseparable if and only if $F>{1\over N}$.

Now, to distill the considered states we need to generalize quantum XOR
gate \cite{Deutsch}. The N-dimensional counterpart of the latter we choose
to be
\begin{equation}
U_{XOR^N}|k\rangle |l\rangle = |k\rangle  |l\oplus k\rangle
\end{equation}
where $k \oplus l= (k+l)\mathop{\rm mod} N $.
The $|k\rangle$ and  $|l\rangle $ states describe the source and target
systems respectively.
Now the protocol is analogous to that in Ref. \cite{Bennett1}.
\begin{enumerate}
\item Two input pairs are twirled i.e. each of them is subjected to
random bilateral rotation of type $U \otimes U^*$
\item The pairs are subjected to
the transformation $U_{XOR^N}\otimes U_{XOR^N}$.
\item The target pair is measured in the basis $| i \rangle \otimes |j \rangle$.
\item If the outcomes are equal, the source pair is kept,
 otherwise it is discarded.
\end{enumerate}
If the outcomes were identical, then  twirling the resulting source pair, we
obtain it
in state $\varrho_{\alpha'}$ where $\alpha'$ satisfies equation
\begin{equation}
\alpha'(\alpha)=\alpha{(N(N+1) - 2)\alpha +2 \over (N+1) (1+(N-1)\alpha^2)}
\end{equation}
The above function is increasing and continuous in total range
$\alpha\in({1 \over N +1}, 1)$.
Hence, as in Ref. \cite{Bennett1}, the fraction increases if the initial fraction was
greater than $1/N$. Then to obtain a nonzero asymptotic yield of distilled pure
entanglement, one is to follow the above protocol to obtain some high fidelity
F and then project locally the resulting state onto two-dimensional spaces.
For F high enough the resulting states on $2\times2$ system can be distilled
by e.g. breeding protocol (see \cite{Bennett1}). If needed  they can be
converted again into states of type $P_{+}$ using technique introduced in
Ref. \cite{conc}.

To summarize, given a large amount of pairs of particles in a state
which violates the condition (\ref{kryt}) (or (\ref{kryt2})), one needs first
to apply the filtering procedure given by operator $A$, and then subject
the particle which passed the filter to the recurrence protocol described
above. Note that the operator $A$, if is to describe the process of filtering
(or a part of generalized measurement), should be properly normalized,
so that $\|A\|\leq1$.

Note that the present results allow for simple, independent
proof of the fact \cite{Rains} that the tensor product of K pairs of
two spin--${1 \over 2}$
Bell diagonal states $\mathop{\otimes}\limits_K \varrho_B$, each
with fidelity $F\leq{1 \over 2}$ can not be transformed
into a state  of $N\times N$ system
with  $F' > {1 \over N}$ by means of separable superoperators
\cite{Knight,Rains} which are defined as
$\Lambda (\varrho)=\sum_{i} A_{i} \otimes B_{i} \varrho
A_{i}^{\dagger} \otimes B_{i}^{\dagger}$.
Indeed,  if a two spin--${1 \over 2}$
Bell diagonal state  $\varrho_{B}$
has $F\leq{1 \over 2}$ then it is separable state \cite{huge,inf}.
On the other hand,  any  state of $N\times N$ system,
say $\sigma_{N}$,
with  $F > {1 \over N}$ is inseparable as it can be
$U \otimes U^{*}$ twirled to the state
(\ref{prod}) with $F>{1 \over N}$ which we have shown to be inseparable.
But no separable state (in particular the one
$\mathop{\otimes}\limits_K \varrho_B$ constructed from
separable states $\varrho_B$) can not be transformed by separable
operations into the inseparable state $\sigma_{N}$.

\section{Examples}
\label{ilustracja}
Here we will illustrate the criterion and the first stage of our
distillation protocol. For this purpose consider the following unitary
embedding the Hilbert space $C^N$ into $C^N\otimes C^N$ \cite{Barnett}
\begin{equation}
|i\rangle\rightarrow |i\rangle \otimes |i \rangle
\end{equation}
By means of this transformation we can ascribe to any state $\varrho^N$
on $C^N$ a state $\varrho^{Ne}$ acting on $C^N\otimes C^N$. For example if
$N=3$ and $\varrho^N$ is given by
\begin{equation}
\varrho^N=
\left[
\begin{array}{ccc}
\varrho_{11}&\varrho_{12}&\varrho_{13}\cr
\varrho_{21}&\varrho_{22}&\varrho_{23}\cr
\varrho_{31}&\varrho_{32}&\varrho_{33}\cr
\end{array}
\right]
\end{equation}
then
\begin{equation}
\varrho^{Ne}=
\left[
\begin{array}{ccccccccc}
\varrho_{11}&0&0&0&\varrho_{12}&0&0&0&\varrho_{13}\\
0&0&0& 0&0&0& 0&0&0\\
0&0&0& 0&0&0& 0&0&0\\
0&0&0& 0&0&0& 0&0&0\\
\varrho_{21}&0&0&0&\varrho_{22}&0&0&0&\varrho_{23}\\
0&0&0& 0&0&0& 0&0&0\\
0&0&0& 0&0&0& 0&0&0\\
0&0&0& 0&0&0& 0&0&0\\
\varrho_{31}&0&0&0&\varrho_{32}&0&0&0&\varrho_{33}\\
\end{array}
\right]
\end{equation}
The reductions of $\varrho^{Ne}$ are both  equal to the state $\varrho^N$ with
off-diagonal terms different from zero. That the state $\varrho^{Ne}$
is inseparable if and only if $\varrho^N$ is not diagonal
can be viewed in different ways. First, the state
$\varrho^{Ne}$ with some off-diagonal elements different from zero
violates the $\infty$ entropy inequality as
$\|\varrho^{Ne}\|=\|\varrho^N\|>\max_j\{\varrho_{jj}\}=\|\varrho^{Ne}_X\|$,
where $X=A$ or $B$ (of course if $\varrho^N$ is diagonal
then $\varrho^{eN}$ is trivially separable).
On the other hand, we can apply the  Peres
criterion. 
However, the two criteria do not say whether and
how the state can be distilled. Then let us apply the reduction criterion.
Here (e.g. for $N=3$) we have
\begin{equation}
\varrho^{Ne}_A\otimes I-\varrho^{Ne}=
\left[
\begin{array}{ccccccccc}
0&0&0&0&-\varrho_{12}&0&0&0&-\varrho_{13}\\
0&\varrho_{11}&0& 0&0&0& 0&0&0\\
0&0&\varrho_{11}& 0&0&0& 0&0&0\\
0&0&0& \varrho_{22}&0&0& 0&0&0\\
-\varrho_{21}&0&0&0&0&0&0&0&-\varrho_{23}\\
0&0&0& 0&0&\varrho_{22}& 0&0&0\\
0&0&0& 0&0&0& \varrho_{33}&0&0\\
0&0&0& 0&0&0& 0&\varrho_{33}&0\\
-\varrho_{31}&0&0&0&-\varrho_{32}&0&0&0&0\\
\end{array}
\right]
\end{equation}
Hence, if only the state $\varrho^{Ne}$ is inseparable then it violates the
criterion. Then we can apply the distillation criterion calculating the
eigenvector corresponding the suitable negative eigenvalue, subjecting the
state to the appropriate filter and performing then the recurrence protocol.
However, it can be checked that for $N=3$ this state has already
the fidelity greater than ${1 \over 3}$, hence it can be distilled
without the filtering step.

Consider now the second, more explicit example. Let $P^{3}_{+}$ denote the triplet  
 state (\ref{tryplet}) with N=3 and let
$P_{ij} = |i \rangle \langle i| \otimes |j \rangle  \langle j|$
\begin{equation}
\sigma=pP^3_{+} + (1-p)P_{12} , \\\ p \leq {1 \over 3}.
\label{nstan}
\end{equation}
It can be proved that fully entangled fraction $f$ of this state is not greater
then $1 \over 3$. For this purpose consider the overlap of the
$U_A\otimes U_B$ transformation of the state $P_{+}$ with an
arbitrary pure state $P_{\Phi}$,
$\Phi=\sum_{i,j=1}^{N}  a_{ij} |i \rangle |j \rangle $.
We obtain that
\begin{equation}
Tr(P_{\Phi}U_A \otimes U_B P_{+}U_A^{\dagger} \otimes U_B^{\dagger})=
Tr(P_{\Phi}I \otimes U_B U_A^{\dagger}P_{+}U_A U_B^{\dagger})=
|Tr(A_{\Phi} U_B U_A^{\dagger})|^2,
\end{equation}
where as in sec. \ref{protokol} the matrix
elements of $A_\Phi$ are $\{ A_{\Phi} \}{}_{ij} = \sqrt N a_{ij}$.
Straightforward computation analogous to the one performed
in Ref. \cite{tel} leads us to the following
formula on fully entangled fraction of pure state $\Phi$
\begin{equation}
f(P_{\Phi})=[ Tr(\sqrt{(A_{\Phi}A_{\Phi}^{\dagger}}) ] ^2=
{1 \over N}(\sum_{i=1}^{N}c_i)^2
\end{equation}
where $c_i$ are Schmidt decomposition (hence positive)
coefficients of the state $\Phi$.
From the above formula it follows that in our case the
fully entangled fraction
of product pure state can not be greater than ${1 \over 3}$.
As we assumed that the probability p is also not greater than this number
we obtain immediately
that fully entangled fraction of the state satisfies
$f(\sigma) \leq {1 \over 3}$.
Now we can apply the prescription given in section \ref{rodzina}.
According to (\ref{nstan})
we have the matrix $\sigma_A \otimes I - \sigma$ of the form
\begin{equation}
\sigma_A\otimes I-\sigma=
\left[
\begin{array}{ccccccccc}
1-p&0&0&0&-{p \over3}&0&0&0&-{p \over3}\\
0&{p \over 3}&0& 0&0&0& 0&0&0\\
0&0&1+{2 \over 3}p & 0&0&0& 0&0&0\\
0&0&0&{p \over 3}&0&0& 0&0&0\\
-{p \over3}&0&0&0&0&0&0&0&-{p \over3}\\
0&0&0& 0&0&{p \over 3}& 0&0&0\\
0&0&0& 0&0&0&{p \over 3} &0&0\\
0&0&0& 0&0&0& 0&{p \over 3} &0\\
-{p \over3}&0&0&0&-{p \over3}&0&0&0&0\\
\end{array}
\right]
\end{equation}
This matrix has negative eigenvalue
$\lambda=
{1 \over 2}( 1 - {4 \over 3}p - \sqrt{1 - {4 \over 3}p + {4 \over 3}p^2})$
with the corresponding eigenvector
\begin{equation}
{1 \over \sqrt{1+2y^2}}(|1 \rangle | 1 \rangle + y|2 \rangle |2 \rangle
+ y |3 \rangle |3 \rangle), \ \ y={1 \over 4p}
(3 - 10p + 3\sqrt{1 - {4 \over 3}p + {4 \over 3}p^2}).
\end{equation}
According to the section \ref{rodzina}, in order to distill some entanglement
from the state we can apply the local filter
\begin{equation}
A=\sqrt{3 \over 1 +2y^2}
\left[
\begin{array}{ccc}
1&0&0\\
0&y&0\\
0&0&y\\
\end{array}
\right]
\label{filtr3}.
\end{equation}
Then we obtain the new state
\begin{equation}
\sigma'={1 + 2y^2 \over 3 - 2p + 2py^2} (p P_{+}+{3(1-p) \over 1 + 2y^2}P_{12}) =
p'P_{+} + (1 - p')P_{12}.
\end{equation}
From the  previous results we know that the new state must have fidelity
greater than ${1 \over 3}$. To see it in this particular case
it suffices
only to show that ${p' \over (1 - p')} > {1 \over 2}$.
This inequality can be transformed to the form :
\begin{equation}
3 - 14p + 22p^2 + (3 -10p)\sqrt{1 - {4 \over 3}p  + {4 \over 3}p^2}>0
\end{equation}
Using the fact that $p\leq {1 \over 3}$, the last term in this formula
can be estimated from below by ${1 \over 3}$. This leads to the inequality
which can be checked directly.
Thus in the process of filtering the input state with the fidelity less
than ${1 \over 3}$ has been transformed into the state  with
$F$ strictly greater than
${1 \over 3}$. Then the protocol using generalised $XOR$ operations
described in section \ref{rodzina} can be applied.
Note that the result of the procedure is independent of the
choice of normalisation of the filter. Thus we can choose the best possible
normalisation multiplying the matrix  (\ref{filtr3}) by the constant
to transform the largest number of  diagonal elements into identities.
It results in optimal filter
\begin{equation}
A=\left[
\begin{array}{ccc}
\sqrt{1 +2y^2 \over 3y}&0&0\\
0&1&0\\
0&0&1\\
\end{array}
\right]
\label{filtr4}.
\end{equation}
Then the probability of successful outcome (after which we can perform
the $XOR$ step) is $q={3 - 2p + 2py^2 \over 3y}$.

\section{Discussion and conclusion}
We have introduced a separability criterion relating the structures of 
total state of the system and its reductions. The criterion (called 
reduction one) has been generated by means of some positive map. 
Subsequently, we have shown that any state violating the reduction criterion
is distillable.
Now, in further investigation leading to the solution of the problem whether
any  state violating Peres condition can be distilled it suffices to
restrict to the states
which satisfy the criterion. On the other hand we have determined limit for
use of a class of protocols i.e. the ones consisting of two steps:
one-side,
single-pair filtering and any procedure which can only distill
the states with fully entangled fraction greater than $1\over N$.
Indeed, if a state can be
distilled by such a protocol, then filtering must increase the fraction to the
larger value than $1\over N$, hence the state violates the criterion.

It is worth to note that to prove that any state violating the reduction
criterion can be distilled   the main task was to distill inseparable
$U\otimes U^*$ invariant states. In a similar way it can be shown that
to be able to distill all the
states violating the partial transposition criterion one needs only to provide
a protocol of distillation of the inseparable $U\otimes U$ invariant states
(Werner states \cite{Werner}). This combined with filtering will produce the
desired result. Up till now, we know how to distill only part of the
Werner states (this can be achieved by using Popescu  result \cite{Pop}),
however the other part cannot be distilled by known methods.

The present criterion may be exploited
together with  two-side filtering and it cannot be excluded that it might
allow to distill states which do not violate it at the beginning.
Then it is interesting to characterize the class of states which initially
do not violate the criterion, but do it if subjected to a one-side filtering.
It is remarkable, that all the states violating the criterion, or violating
it after local transformations are nonlocal, which follows
from consideration of the distillation process in the context of
sequential hidden variable model \cite{sekw}.

The reduction criterion divides the set of inseparable states into two classes
of states:
the ones that violate it and the ones that satisfy it. It seems that the
former ones possess analogous properties to the inseparable two-qubit
states. In particular, there is a hope that methods  which have
been successfully applied to the two-qubit states (or one-qubit quantum
channels) like e.g. weight enumerator techniques \cite{Rains2,Rains}
will also work well
for states violating the reduction criterion (or
corresponding noisy channels). Then the latter states
 could be called two-qubit-like states. In contrast, the inseparable
states satisfying the criterion are supposed to exhibit features which never
occur in the two-qubit case. Then, to deal with these states, completely
new methods must be worked out. An example of such states are Werner states,
for which no direct generalization of two-qubit methods leads to
distillation.

Finally,  note that both the positive maps applied so far in investigations of
separability have some physical sense. The transposition means changing
the direction of time \cite{time}. The present positive
map if applied to a part of a compound system indicates a nonzero content
of pure entanglement in the  state of the system.
Then we believe that further investigation of inseparability by
means of positive
maps could allow us not only to characterize the set of separable states,
but also to reveal a possible physical meaning of maps which are positive but
not completely positive.

We are indebted to A. Kossakowski and A. Uhlmann for discussion on
positive maps, R. Horodecki for many helpful comments and
E. Rains for allowing us to incorporate his proof concerning collective
application of reduction criterion and for helpful comments.
We are also grateful to  N. J. Cerf and R. M. Gingrich
for information on their numerical results, which contributed to
removing an error that appeared in an earlier version of this paper.
P.H. acknowledges the 1997 Elsag-Bailey -- I.S.I.
Foundation research meeting on quantum computation.
The work is supported by Polish Committee for Scientific Research,
Contract No. 2 P03B 024 12 and by Foundation for Polish Science.

\vskip1cm

\begin{appendix}
\section{}
Here we will prove that the positive map $\Lambda$ given by eq. (\ref{map}) is
decomposable i.e. it can be written in the form \cite{Woronowicz}
\begin{equation}
\Lambda= \Lambda_1^{CP}+T\Lambda_2^{CP},
\end{equation}
where $\Lambda_i^{CP}$ are CP maps and $T$ is transposition.
 In fact, we will see that the map is trivially
decomposable i.e. it is of the form $\Lambda=T\Lambda^{CP}$.
To prove the above we need the lemma establishing one-to-one correspondence
between CP maps $\Lambda: M_N\rightarrow M_N$  and
positive matrices (operators) belonging to  tensor product $M_N\otimes M_N$ 
(this is analogous to the fact that positive maps are equivalent to
the matrices in $M_N\otimes M_N$ which are positive on product vectors
\cite{Jamiolkowski,sep}).

{\it Lemma.} A linear map $\Lambda: M_N\rightarrow M_N$ is completely
positive if and only if the operator $D\in M_N\otimes M_N$ given
by
\begin{equation}
D=(I\otimes \Lambda) P_+
\label{oper}
\end{equation}
is positive (here $P_+$ is given by eq. (\ref{tryplet})).

{\it Proof.} If $\Lambda$ is CP then by the very definition of the CP map,
the operator $D$ is positive. Conversely, suppose  that
the operator $D$ is positive. Then it can be written by means of its
spectral decomposition
\begin{equation}
D=\sum_i\lambda_i|\psi_i\rangle\langle\psi_i|\nonumber
\end{equation}
with nonnegative eigenvalues $\lambda_i$. Taking $V_i$ such that
$I\otimes V_i|\psi_+\rangle=|\psi_i\rangle$ (see sec. \ref{protokol}),
we obtain
\begin{equation}
D=\sum_i\lambda_i I\otimes V_i P_+ I\otimes V_i^\dagger\nonumber
\end{equation}
Comparing this formula with eq. (\ref{oper})
and noting that $\Lambda$  is uniquely determined by this equation,
we obtain  that it is given by
\begin{equation}
\Lambda(\sigma)=\sum_i W_i \sigma W_i^\dagger
\nonumber
\end{equation}
where $W_i=\sqrt{\lambda_i} V_i$. However this is the general form of
completely positive maps \cite{Kraus}.
This ends the proof of the lemma.

{\it Remark.} The lemma also holds  for $\Lambda:M_N\rightarrow M_K$ with
$N\not= K$. Then the $P_+$ belongs to $M_N\otimes M_N$ and
the operator $D$ belongs to $M_N\otimes M_K$.

Consider now the map of interest  given by $\Lambda(\sigma)=I{\rm Tr}\sigma-
\sigma$.
The corresponding operator $D$ (by eq. (\ref{kryt}) is given by
\begin{equation}
D= (P_+)_A\otimes I -P_+\nonumber
\end{equation}
where $(P_+)_A$ is the reduction of the state $P_+$ so that $(P_+)_A={1\over
N}I$. Consider now the partial transposition of $B$. It can be checked that
$D^{T_B}$ is of the form
\begin{equation}
D^{T_B}={1\over N} (I\otimes I - V)\nonumber
\end{equation}
where $V$ is the operator \cite{Werner} defined by
$V\psi\otimes\phi=\phi\otimes\psi$ for any vectors $\phi,\psi\in C^N\otimes C^N$.
As $V^2=I\otimes I$ we obtain that $V$ has eigenvalues $\pm1$ so that
$I\otimes I -V$ is a positive operator. Thus we see that $D^{T_B}$ is a
positive operator.  However we have $D^{T_B}=(I\otimes T\Lambda)P_+$.
Then by the lemma the map $\Gamma=T\Lambda$ is CP. Consequently,
we obtain
\begin{equation}
\Lambda=T\Gamma\nonumber
\end{equation}
which ends the proof. Of course, $\Lambda$ can be also written as
$\Lambda=\Gamma' T$ with completely positive $\Gamma'$. Indeed, as
$\Gamma$ is CP, then it is of the form
$\Gamma(\sigma)=\sum_iV_i\sigma V^\dagger_i$. Hence
\begin{equation}
T(\Gamma(\sigma))=\sum_i(V_i\sigma V^\dagger_i)^T=\sum_i(V_i^T)^\dagger
\sigma^T V_i^T=\sum_i\tilde V_i \sigma^T\tilde V^\dagger_i
\equiv\Gamma'(T(\sigma))
\end{equation}
with $\tilde V_i=(V_i^T)^\dagger$. Thus $\Gamma'$ is completely positive.
\end{appendix}


\begin{references}
\bibitem[*]{poczta1} E-mail address: michalh@iftia.univ.gda.pl
\bibitem[**]{poczta2} E-mail address: pawel@mifgate.pg.gda.pl
\bibitem{EPR}
A. Einstein, B. Podolsky and N. Rosen, Phys. Rev.  {\bf 47}, 777  (1935);
E. Schr\"odinger,  Nat\"urwissenschaften {\bf 23}, 807 (1935); J. S. Bell
Physics (N.Y.) {\bf 1}, 195  (1964).
\bibitem{Ekert}
A. Ekert, Phys. Rev. Lett. {\bf 67}, 661 (1991).
\bibitem{geste}
C. H. Bennett  and S. J. Wiesner, Phys. Rev. Lett. {\bf 69}, 2881 (1992).
\bibitem{Bennett_tel}
C. Bennett, G. Brassard, C. Crepeau, R. Jozsa, A. Peres and W. K. Wootters,
Phys. Rev. Lett.  {\bf 70}, 1895  (1993).
\bibitem{huge}
C. H. Bennett, D. P. Di Vincenzo, J. Smolin and
W. K. Wootters, Phys. Rev. A {\bf 54},	3814 (1997).
\bibitem{compl}
R. Cleve and H. Burhman, Phys. Rev. A, {\bf 56}, 1201 (1997).
\bibitem{geste_exp}
K. Mattle, H. Weinfurter, P. Kwiat and A. Zeilinger, Phys. Rev. Lett.
{\bf 76}, 4656 (1996).
\bibitem{Zeilinger}
D. Bouwmeester, J.-W. Pan, K. Mattle, M. Elbl, H. Weinfurter and A. Zeilinger,
Nature (London) {\bf 390}, 575 (1997); D. Boschi, S. Brance, F. de Martini, L.
Hardy and S. Popescu, Phys. Rev. Lett. {\bf 80}, 1121 (1998).
\bibitem{Bohm}
D. Bohm, Phys. Rev. {\bf 85}, 166 (1952).
\bibitem{Bennett1}
C. H. Bennett, G. Brassard, S. Popescu, B. Schumacher, J. Smolin and
W. K. Wootters, Phys. Rev. Lett. {\bf 76}, 722 (1996).
\bibitem{mixed}
Throughout the paper we will identify states with their density matrices.
\bibitem{Werner}
R. F. Werner, Phys. Rev. A {\bf 40}, 4277 (1989).
\bibitem{konieczne}
In this paper we will refer to a necessary condition for separability
(i.e. the condition which must be satisfied by any separable state)
as to separability condition or separability criterion.
\bibitem{bell}
R. Horodecki, P. Horodecki and M. Horodecki, Phys. Lett. A {\bf 200}, 340
 (1995).
\bibitem{red}
R. Horodecki and P. Horodecki, Phys. Lett. A {\bf 194}, 147 (1994).
\bibitem{renyi}
R. Horodecki, P. Horodecki, and M. Horodecki, Phys. Lett. A {\bf 210}, 377
(1996).
\bibitem{inf}
R. Horodecki and M. Horodecki, Phys. Rev. A
{\bf 54}, 1836 (1996).
\bibitem{Peres}
A. Peres, Phys. Rev. Lett. {\bf 76}, 1413 (1996);
\bibitem{sep}
M. Horodecki, P. Horodecki and R. Horodecki, Phys. Lett. A
{\bf 223}, 1 (1996).
\bibitem{qubit}
throughout this paper the $N\times M$ system means the system described by the
Hilbert space $C^N\otimes C^M$ so that e.g. $2\times2$ system corresponds to
the two spin-$1\over2$ or, in general, two two-level (two-qubit) one.
\bibitem{Paw}
P. Horodecki, Phys. Lett. A {\bf 232}, 333 (1997).
\bibitem{Gisin}
N. Gisin, Phys. Lett. A {\bf 210}, 151 (1996).
\bibitem{conc}
C. H. Bennett, H. J. Bernstein, S. Popescu and B. Schumacher, Phys. Rev. A
{\bf 53}, 2046 (1996).
\bibitem{pur}
M. Horodecki,  P. Horodecki and R. Horodecki, Phys. Rev. Lett.
{\bf 78}, 574 (1997).
\bibitem{distbis}
M. Horodecki, P. Horodecki and R. Horodecki, Report No. quant-ph/9801069,
Phys. Rev. Lett. (in press).
\bibitem{Lindblad}
G. Lindblad, Commun. Math. Phys. {\bf 40}, 147 (1975), W. F.  Stinespring
Proc. Am. Math. Soc. {\bf 26}, 211 (1965).
\bibitem{Kraus}
K. Kraus, {\it States, Effects and Operations: Fundamental Notions of
Quantum Theory}, Wiley, New York, 1991.
\bibitem{time}
P. Busch and J. Lahti, Found. Phys. {\bf 20}, 1429 (1990);
A. Sanpera, R. Tarrach and G. Vidal, Quantum separability,
time reversal and canonical decomposition, quant-ph/9707041.
\bibitem{Woronowicz}
S. L. Woronowicz, Rep. Math. Phys. {\bf 10}, 165  (1976).
\bibitem{Kossakowski}
A. Kossakowski, private communication.
\bibitem{ac}
After completion of the first version of this
manuscript we received an information that this condition was independently
derived and discussed
by N. J. Cerf, C. Adami and R. M. Gingrich (quant-ph/9710001).
\bibitem{Gin}
This is in agreement with the numerical evidence obtained by
N. Cerf and R. Gingrich (private communication).
\bibitem{Popescu}
S. Popescu,  Phys. Rev. Lett.  {\bf 72},  779 (1994).
\bibitem{Rains_priv}
E. Rains, (private communication).
\bibitem{Nielsen}
B. Schumacher and M. A. Nielsen, Phys. Rev. A {\bf 54}, 2629 (1996).
\bibitem{Jozsa}
R. Jozsa, J. Mod. Opt. {\bf 41}, 2315 (1994).
\bibitem{Shor}
P. W. Shor and J. A. Smolin, Report No. quant-ph/9604006;
D. DiVincenzo, P. W. Shor and J. A. Smolin,
Phys. Rev. A {\bf 57}, 830 (1998).
\bibitem{Deutsch}
D. Deutsch, Proc. R. Soc. London A {\bf 425}, 73 (1989).
\bibitem{Rains}
E. M. Rains, Report No. quant-ph/9707002.
\bibitem{Knight}
V. Vedral, M. B. Plenio, M. A. Rippin and P. L. Knight
Phys. Rev. Lett. {\bf 78} (1997) 2275.
\bibitem{Barnett}
S. M. Barnett and S. J. D. Phoenix, Phys. Rev. A {\bf 44}, 535 (1991).
\bibitem{tel}
R. Horodecki, M. Horodecki and P. Horodecki, Phys. Lett. A  {\bf 222}, 21
(1996).
\bibitem{sekw}
N. D. Mermin, Hidden quantum non-locality (1995). Preprint, Lecture given in
Bielefeld at the conference ``Quantum mechanics without observer'';
S. Popescu, More powerful tests of nonlocality
by sequences of measurements, in ``The dilemma of Einstein, Podolsky and 
Rosen - 60 years after'', proceedings of ``60 years of EPR'' conference,
March 1995, Haifa Israel, IOP Publishing USA; see also
S. Teufel, K. Brendl, D. D\"urr, S. Goldstein, N. Zangh\'{\i}
Phys. Rev. A {\bf 46}, 1217 (1997);
M. \.Zukowski, R. Horodecki, M. Horodecki and P. Horodecki,
Generalized measurements and local realism, quant-ph/9608035.
\bibitem{Pop}
S. Popescu, Phys. Rev. Lett. {\bf 74}, 2619 (1995).
\bibitem{Rains2}
P. Shor and R. Laflamme, Phys. Rev. Lett. {\bf 78}, 1600 (1997); E. Rains
Report No. quant-ph/961001, quant-ph/9612015.
\bibitem{Jamiolkowski}
A. Jamio\l{}kowski, Rep. Math. Phys.,  3 (1972) 275.
\end{references}
\end{document}